# Variations of Particle Motion in the Van Allen Belts

Ariana Bussio, *University of California Los Angeles 2013, Howard Community College*
Mentored by: Alex Barr, Ph.D

## Abstract

*Earth's magnetic field traps charged particles from the solar wind in the Van Allen belts. These trapped ions execute three types of motion: helical motion around magnetic field lines, longitudinal motion between North and South mirror points, and latitudinal drift motion around Earth. Each is associated with an adiabatic invariant, which expresses restrictions on each type of motion. This study examines the effect on proton motion and the three adiabatic invariants when the strength of Earth's magnetic moment varies with time. Using a Python 3 code developed in Google Colaboratory, each invariant was calculated and graphed for a static magnetic field, for magnetic fields that varied with periods shorter than the invariant period, and for magnetic fields that varied with periods longer than the invariant period. The results demonstrate that if Earth's magnetic moment changes quickly enough with time, the adiabatic invariant is no longer constant, allowing a wider variety of ion movement. Further study of these invariants and how they change in time-varying magnetic fields could help predict ion motion and changes to the magnetosphere during events that affect space weather, like solar storms or pole reversals.*

## Introduction

Earth's magnetic field plays an important role in deflecting radiation from solar wind and protecting Earth's atmosphere. The interaction of Earth's magnetic field and the Sun's solar wind causes "space weather" like geomagnetic storms and aurora phenomena. Fluctuations and extreme interactions caused by events like solar flares or Coronal Mass Ejections cause often unpredicted changes to Earth's protective magnetic field and damage to equipment both in orbit and on the ground. These events can also pose threats to space travel, exposing astronauts to harmful radiation [1].

Earth's Van Allen radiation belts are zones of the magnetosphere which capture and retain charged particles. Physical evidence for the Van Allen belts began interesting researchers in the late 1950s with data from *Sputnik II* and *Explorer I*, and it became clear in the 50s and 60s that ions are trapped in belts by Earth's magnetic field [2]. Two Van Allen belts are always present: an inner belt consisting mostly of protons with energies from 10 MeV to 50 MeV [3] and an outer belt consisting mostly of electrons with energies from .04 MeV to around 7 MeV [4]. The regions are separated by a "slot." A third temporary belt produced by major solar events and their subsequent geomagnetic storms has recently been observed in this slot between the two belts [2].

The motion of ions in the Van Allen belts is restricted by three adiabatic invariants, each of which restricts a type of motion: "helical motion," the spiral motion an ion engages in around a magnetic field line; "bounce motion," where an ion travels along a field line between North and South magnetic poles; and "drift motion," where the ion travels equatorially (perpendicular to bounce motion) around Earth in a shell [5]. However, these invariants are defined for a static magnetic field. Earth's magnetic field varies during geomagnetic storms and during pole reversals, diminishing the constraints associated with these invariants and allowing a wider range of ion motion. This paper examines each adiabatic invariant in a static magnetic field and in magnetic fields that vary periodically with time in order to help identify the role of adiabatic invariants in constraining ion motion during periods of calm and during space weather events.

Particle Motion

A Python 3 code was developed in Google Colaboratory in order to graph and examine the motion of ions and the adiabatic invariants that restrict motion. Through the fourth order Runge-Kutta method of numerical approximation, the force, acceleration, velocity, and position of each particle was calculated. Earth's magnetic field was modeled as a magnetic dipole:

$$\boldsymbol{B} = \frac{\mu_0}{4\pi} r^3 [3(\boldsymbol{M} \cdot \hat{\boldsymbol{r}})\hat{\boldsymbol{r}} - \boldsymbol{M}] \qquad (1)$$

where $\mu_0$ is the permeability of free space, **M** is the vector of Earth's magnetic moment and **r** is the ion's position vector from Earth's center [6]. The acceleration of the protons was calculated using proton mass and the Lorentz equation for force, which states that the force acting on a particle is equal to the cross product of the product of its charge and velocity (q**v**) and the local magnetic field (**B**) [7]. The effects of the electric field in Earth's magnetosphere are small enough that the electric field can be modeled as zero in an examination of adiabatic invariants and their restrictions on particle motion [5].

$$\boldsymbol{F} = q\boldsymbol{v} \times \boldsymbol{B} \qquad (2)$$

The kinetic energy of the ions is expressed by the relativistic equation:

$$KE = mc^2 \left( \frac{1}{\sqrt{1 - \frac{v^2}{c^2}}} - 1 \right) \qquad (3)$$

This can also be expressed in terms of the relativistic factor $\gamma$:

$$KE = mc^2(\gamma - 1) \qquad (4)$$

Adiabatic Invariants

Using the Python 3 code, the three adiabatic invariants which restrict particle motion were calculated and graphed with respect to time. The program was first run with the following initial conditions:

|  | Ion 1 | Ion 2 |
|---|---|---|
| Mass (m) | $1.672621777 \times 10^{-27}$ kg | $1.672621777 \times 10^{-27}$ kg |
| charge (q) | $+1.602176565 \times 10^{-19}$ C | $+1.602176565 \times 10^{-19}$ C |
| initial kinetic energy ($KE_0$) | $q \times 10^7$ J (10 MeV) | $q \times 10^7$ J (10 MeV) |
| initial velocity (**v**) | $\left[0, c\sqrt{1-\frac{1}{\gamma^2}}\sin(\theta), c\sqrt{1-\frac{1}{\gamma^2}}\cos(\theta)\right]$ m/s | $\left[0, c\sqrt{1-\frac{1}{\gamma^2}}\sin(\theta), c\sqrt{1-\frac{1}{\gamma^2}}\cos(\theta)\right]$ m/s |
| initial position (**r**) | $[2R_e, 0, 0]$ m | $[4R_e, 0, 0]$ m |
| angle between **B** and **v** ($\theta$) | $\pi/6$ radians | $\pi/6$ radians |

| Earth Radius ($R_e$) | 6378137 m |
|---|---|
| Equatorial strength of Earth's magnetic dipole field ($B_0$) [8] | $3.07 \times 10^{-5}$ T |
| Strength of Earth's magnetic moment (pz) | $4\pi B_0 R_e^3/\mu_0 = \sim 8 \times 10^{21}$ J/T |
| Earth's magnetic moment (**M**) | pz[0,0,-1] |

**Table 1:** Constants and initial values.

Each invariant can be calculated by integrating

$$\oint (\gamma m \mathbf{v} + q\mathbf{A}) \cdot d\mathbf{l} \tag{5}$$

over one period of the relevant motion, where **A** is the vector potential of the magnetic field [7]. For the first adiabatic invariant, this is one gyro-orbit, estimated at 0.1 seconds [5]. For the second adiabatic invariant, the relevant motion is one half period of bounce motion. These half-periods of bounce motion were individually calculated in the Google Colaboratory program and ranged from 0.5 to 2 seconds. For the third adiabatic invariant, the period of relevant motion is one rotation around Earth [5], each of which were also calculated in the Google Colaboratory program at between 155 and 160 for Ion 1 and between 70 and 77 seconds for Ion 2. Öztürk provides a useful simplification of the integrals used to calculate each adiabatic invariant [9].

The first invariant ($\mu_1$) expresses a conservation of the magnetic moment of a charged particle moving in a circular path. This also implies a conservation of magnetic flux through the region traced by the gyro-orbit while $\mu_1$ is constant [7]. The first invariant can be calculated at each moment in time using $\gamma$, the value of momentum perpendicular to **B** at that time, and the magnitude of **B** at that time [9].

$$\mu_1 = \frac{\gamma^2 m v_\perp^2}{2B} \tag{6}$$

If the first invariant is constant and an ion enters regions of stronger magnetic field, the perpendicular velocity must increase. To conserve energy, the parallel component of velocity must decrease, resulting in a mirror point. The second adiabatic invariant (I) is related to longitudinal motion and the mirror points at each end of the "bounce" arc. The second invariant is calculated by integrating the component of particle momentum parallel to the **B** field over a half period of "bounce" along a field line: for example, from a position directly above the equator (z=0) to the North mirror point and back to the position above the equator (z=0) [9].

$$I = \int_0^{T_{bounce}/2} \frac{v_\parallel^2}{v} dt \tag{7}$$

The third adiabatic invariant ($\Phi$) is related to a particle's drift motion, or its latitudinal movement around Earth, and expresses that the net magnetic flux enclosed by a particle's drift orbit is constant [5]. Therefore, particles are expected to remain in their drift shells and not increase or decrease in average altitude. The net flux within the drift orbit is equal to the net flux on the exterior of the orbit, so flux ($\Phi$) can be calculated as an integral from the particle's equatorial distance ($R_0$) to a point infinitely far away [7].

$$\Phi = \int_{R_0}^{\infty} B_0 \left(\frac{R_E}{r}\right)^3 2\pi r \, dr = 2\pi B_0 \frac{R_E^3}{R_0} \tag{8}$$

In a static **B** field, or in a **B** field that changes slowly with respect to the relevant period of motion, all three quantities ($\mu_1$, I and $\Phi$) will remain approximately constant [7]. When these invariants are constant, the motion associated with each of them is predictable. For example, the radius of an ion's gyro-orbit remains constant and the altitude of the particle's drift shell will remain constant.

This study compares two protons of equal initial kinetic energies, though electrons are more commonly found in the outer Van Allen belt and their energies typically differ from protons in the inner belt, in order to more clearly see differences in particle motion which could not be caused by differences in mass, initial kinetic energy, or charge. Earth's magnetic moment (**M**) was modeled to point directly in the -z direction, toward Earth's magnetic North Pole (geographic South Pole). Öztürk's calculations were used as a guide for initial kinetic energy, $\theta$, and velocity values, though it is important to note that Öztürk's initial velocity calculations contain an error. Values for the invariants will thus vary slightly compared to those in this study [5].

After an initial data set was established with the conditions listed in Table 1, pz was altered to vary sinusoidally with time, making the strength of Earth's magnetic moment (**M**) and Earth's dipole field (**B**) no longer constant, but time dependent as they might be during a solar storm or magnetic pole reversal. While the variation during such an event is likely to be much more complex, this simple model nevertheless serves to illustrate the impact of a magnetic field that varies with time. The time-varying magnetic moment is given by

$$pz = \frac{4\pi B_0 R_E^3}{\mu_0} + A sin(bt) \tag{9}$$

where $A$ was set to $3\times10^{22}$ J/T which was chosen after some trial and error to show effects on particle motion. This resulted in a **B** field which varied by up to 3000 nT. This maximum variation is about 1 to 2 orders of magnitude larger than recorded changes in Earth's magnetic field during previous powerful geomagnetic storms [10]. Similar but less drastic results can be seen with a smaller $A$ value and therefore smaller changes in Earth's magnetic field, however a larger magnitude of variation exaggerates the resulting changes in motion, making them easier to examine. The period of this variation was then adjusted by changing $b$ to examine the effects of relatively "long" or "short" periods of variation on each of the adiabatic invariants and on particle motion.

Each adiabatic invariant was calculated and graphed under several conditions:
a) **B** does not vary with time
b) **B** varies with a period greater than the period of the third adiabatic invariant for Ion 2 ($T_1$=105 s).
c) **B** varies with a period between the periods of the second and third adiabatic invariant for both ions ($T_2$=20 s)
d) **B** varies with a period between the periods of the first and second adiabatic invariant for both ions ($T_3$=0.3 s)
e) **B** varies with a period shorter than the period of the first adiabatic invariant for both ions ($T_4$=0.05 s)

## Results

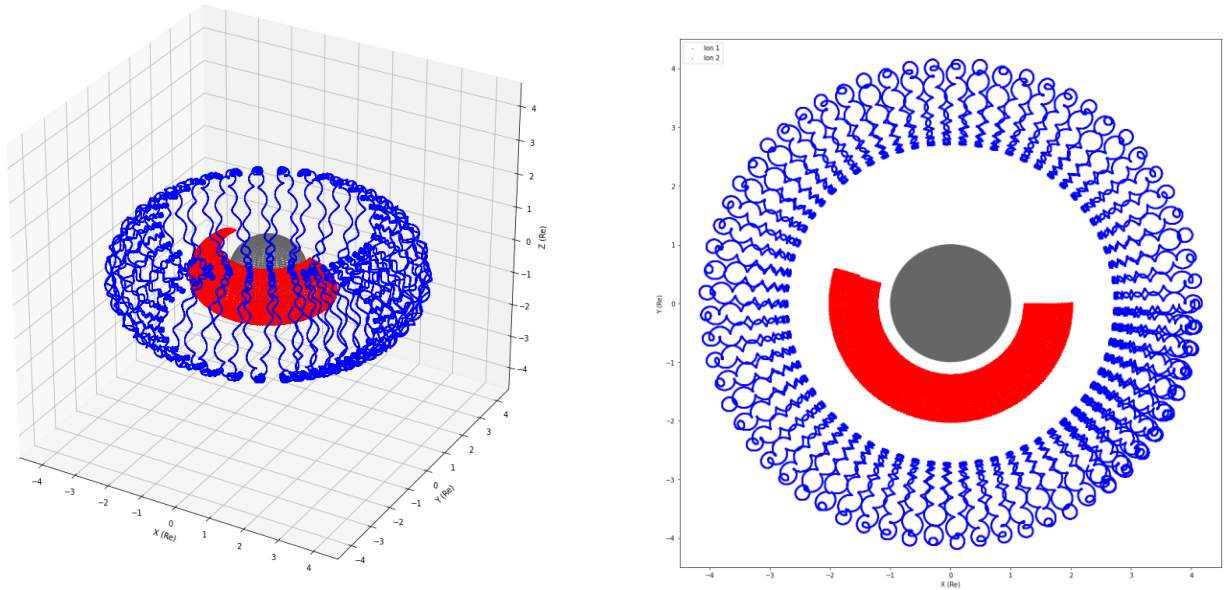

**Figure 1a:** Motion of Ion 1 and Ion 2 over 85 seconds. Three types of motion can be observed: gyromotion around a field line, bounce motion in z and -z directions along field lines, and drift motion around Earth (light grey). Axes range from -4$R_e$ to 4$R_e$ in each dimension.

**Figure 1b:** Motion of Ion 1 and Ion 2 in the x-y plane over 85 seconds, which shows that the shell traced by the proton further from Earth is wider than the shell traced by the proton closer to Earth. Axes range from -4$R_e$ to 4$R_e$ in each dimension.

Figure 1a shows the motion of two protons over 85 seconds. Ion 1 (inner particle) began at an initial position of (2$R_e$,0,0) and Ion 2 (outer particle) began at an initial position of (4$R_e$,0,0). Earth is centered at (0,0,0) with a radius of 6378137 m (1 $R_e$). These positions were chosen to show two distinct particles at two different altitudes (in two different "shells") which are qualitatively similar to the inner and outer Van Allen belts. They are not separated by the physics governing the separation of the two Van Allen belts which are not included in the current model, but are separated simply by their initial assigned position.

First Invariant (μ$_1$)

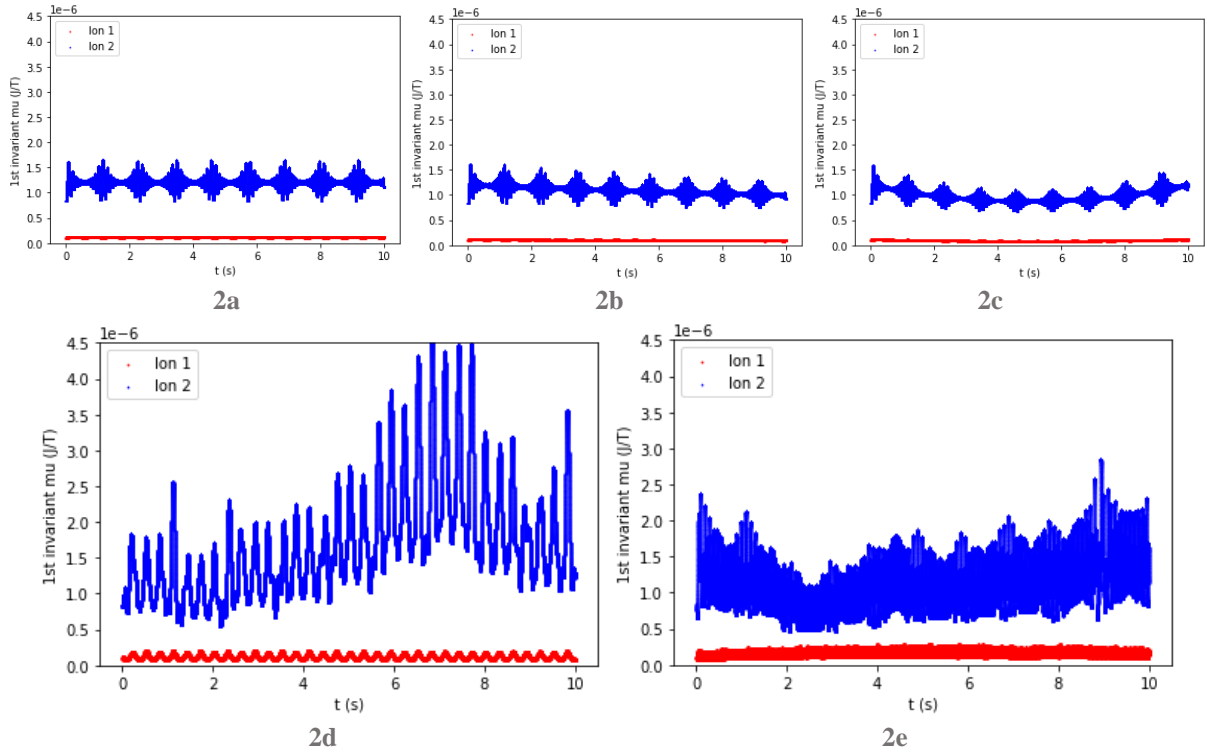

**Figure 2:** First adiabatic invariant of Ion 1 (bottom line) and Ion 2 (top line) over 10 seconds. The period of relevant motion (one gyro-orbit) is about 0.1 seconds. Figures are listed from top left to bottom right.
**Figure 2a:** Constant magnetic moment (**M**). **Figure 2b:** Period of magnetic moment variation: $T_1$=105 s. **Figure 2c:** Period of magnetic moment variation: $T_2$=20 s. **Figure 2d:** Period of magnetic moment variation: $T_3$=0.3 s. **Figure 2e:** Period of magnetic moment variation: $T_4$=0.05 s.

The first invariant was evaluated for each data point. Figure 2 compares first adiabatic invariant values over 10 seconds. Figure 2a shows that μ$_1$ for both ions is constant on average when **M** and **B** do not vary with time. Figures 2b and 2c show that with a period as low as 20 seconds, μ$_1$ is still relatively constant, but does change minutely over time. Figures 2d and 2e show that when the magnetic field varies with shorter periods over time (0.3 s and 0.05 s respectively), μ$_1$ is no longer constant.

Second Invariant (I)

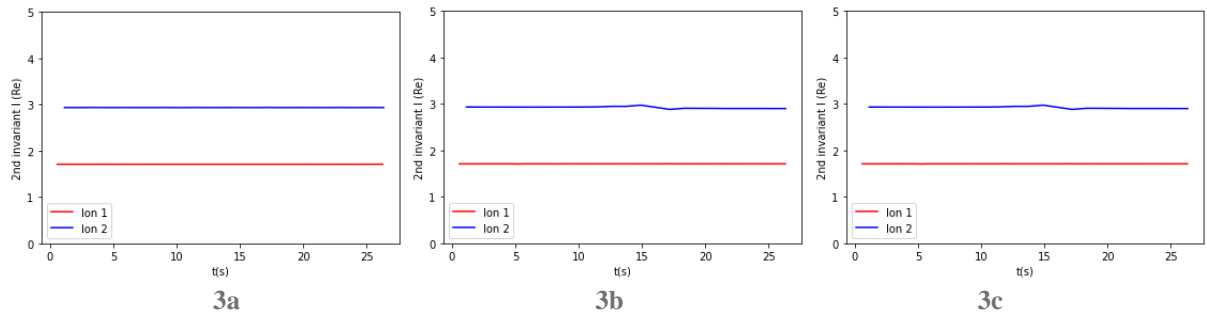

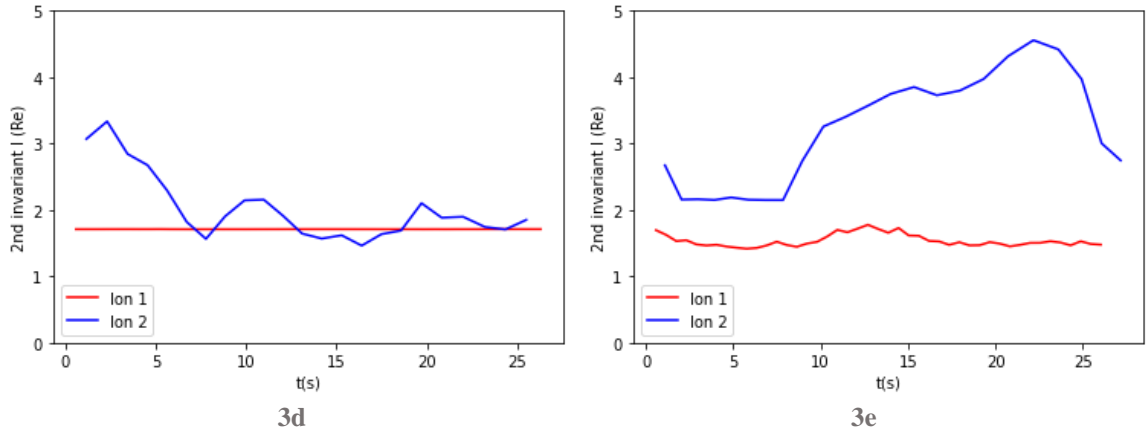

**Figure 3:** Second adiabatic invariant for constant and varying magnetic moments. The period of relevant motion (one half period of bounce) is about 0.5 – 2 seconds. Figures are listed from top left to bottom right.
**Figure 3a:** Constant magnetic moment (**M**). **Figure 3b:** Period of magnetic moment variation: $T_1$=105 s. **Figure 3c:** Period of magnetic moment variation: $T_2$=20 s. **Figure 3d:** Period of magnetic moment variation: $T_3$=0.3 s. **Figure 3e:** Period of magnetic moment variation: $T_4$=0.05 s.

The second invariant was evaluated over each half period of bounce motion. Figure 3 compares the second adiabatic invariant values over 30 seconds. Figures 3b and 3c show that with a period of variation 20 s and above, I is constant. Figure 3c shows that with a period of 0.3 s, the second adiabatic invariant of Ion 1 is constant, while the second adiabatic invariant of Ion 2 shows variations. Figure 3b shows that with a period of 0.05 s, slight variations appear in Ion 1, and larger variations appear in the second adiabatic invariant of Ion 2. One value was calculated for each interval of integration, so there are fewer data points per second for the second adiabatic invariant compared to the first invariant.

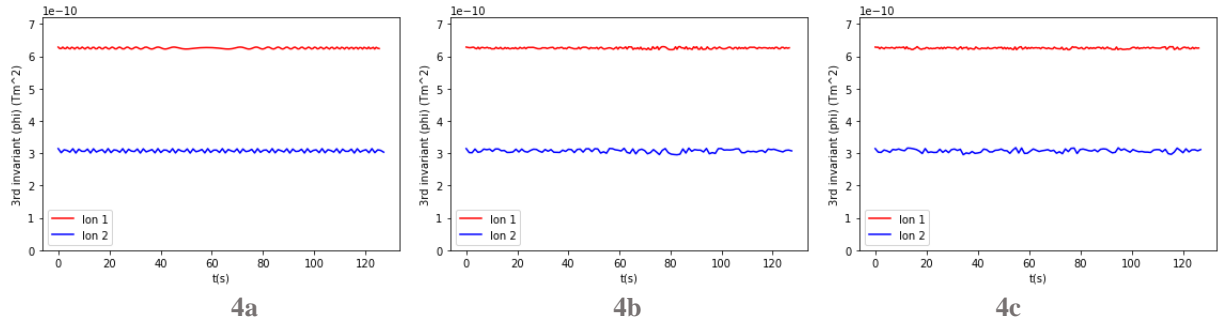

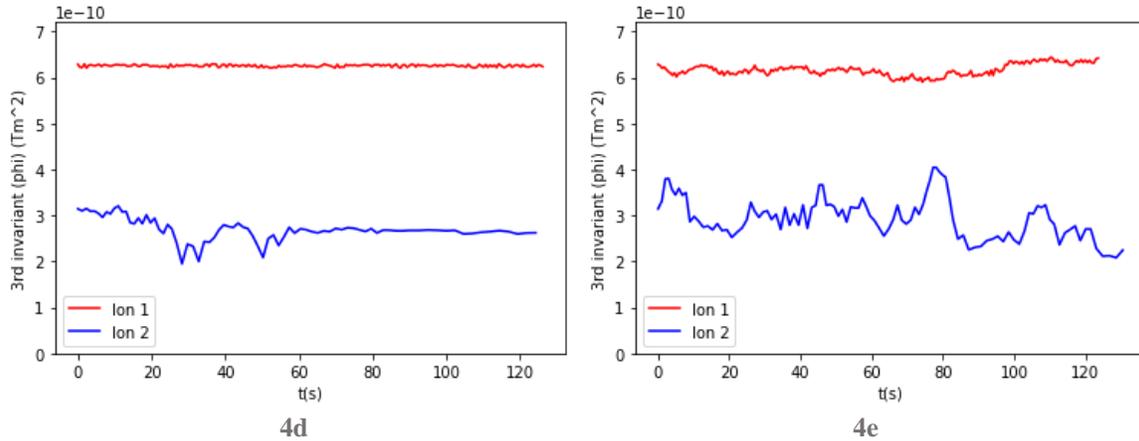

Figure 4: Third adiabatic invariant for constant and varying magnetic moments. The period of relevant motion (one orbit around Earth) is between 70 and 77 seconds for the outer particle and between 155 and 160 seconds for the inner particle. Figures are listed from top left to bottom right.
**Figure 4a:** Constant magnetic moment (**M**). **Figure 4b:** Period of magnetic moment variation: $T_1$=105 s. **Figure 4c:** Period of magnetic moment variation: $T_2$=20 s. **Figure 4d:** Period of magnetic moment variation: $T_3$=0.3 s. **Figure 4e:** Period of magnetic moment variation: $T_4$=0.05 s.

The third adiabatic invariant was calculated using Eq. (8) using the equatorial altitude of each bounce as $R_0$, so the intervals used were also the half periods of bounce motion used in the second invariant. These intervals were used in order to obtain several data points for this shorter simulation where the total run time was less than two orbits around Earth. For a longer simulation, the period of orbit around Earth should be used as the interval and the bounds of integration. Figure 4 compares the third adiabatic invariant values over 125 seconds. There is an increase in invariant fluctuation from 4a (no time variation) to 4b (period of variation of 105 seconds) and these fluctuations in third invariant values continue to increase as period decreases (graphs 4c-4e).

**Conclusions**

The period of variation of the strength of Earth's magnetic moment affects each adiabatic invariant differently. However, each invariant is similar in that as the period continues to decrease, the values vary more and the invariant is less constant. This suggests that particle motion gradually becomes less constrained as the rate at which the strength of the magnetic moment changes, rather than suddenly becoming less regular and predictable at a certain period length.

In general, when the period of variation of Earth's **B** field is greater than the period of motion, the adiabatic invariant remains constant. When the period is below the period of variation, the adiabatic invariants are no longer constant and their values fluctuate. If the period is even lower and the frequency even higher, then the adiabatic invariants vary even more.

The invariant with the longest period of motion is the third, therefore it is the most easily breached. The third invariant limits ions to their drift shells, so it is most likely that ions will drift radially from their shell. The second invariant and the bounce motion it limits is the next most easily breached, while the first invariant requires the highest frequency of magnetic field variation to breach. Measuring the period of magnetic field fluctuation could help predict particle motion by predicting what type of motion will or will not be constrained by the three adiabatic invariants, and at which altitudes the effect will be greatest as the adiabatic invariants of ions at different altitudes varied in different ways.

The strength of Earth's dipole moment and the frequency at which it varies affects the threshold at which the three adiabatic invariants are no longer constant for ions in the Van Allen belts. When **B** is modeled to vary with time, particle motion becomes less regular, especially in the outer Van Allen belt. This suggests that if the frequency of variation is high enough, particle motion could become unpredictable and particles could possibly escape their magnetic field lines and the magnetosphere entirely.

**Further Explanations and Questions**

While most particles in the outer Van Allen Belt are electrons, it was useful to model both ions we examined as protons. Comparing identical particles in two different radiation belts allows conclusions to be drawn about their motion without the uncertainty of added discrepancies between the ions like charge, mass, and kinetic energy. For further study, it would be interesting to compare the movements of a proton in the inner belt and an electron in the outer belt by changing initial mass, charge, and energy values in the program.

Experimenting with the periods of magnetic moment variation could also provide more information on particle motion in changing magnetic fields. For example, setting the period of variation to approach different periods of motion could result in resonance and possibly identify a specific "breaking point" where adiabatic invariants would no longer be useful in calculations or predictions of particle motion. Adjusting particle energy and position could also affect where adiabatic invariants no longer restrict motion. Finding these "breaking point" values may allow us to answer questions like: "How strong would a solar storm have to be to allow 10% of electrons in the outer Van Allen belt to escape?" or "How will Earth's magnetosphere change during a magnetic pole reversal?"

Other important elements to consider for future study are the bowing effect and asymmetry of the magnetosphere created by solar wind and the accelerating effects of waves in the magnetosphere on ions along field lines. Ideally these would be included in the study of particle motion to create more accurate predictive models.

Contact: arianabussio@gmail.com